\documentclass[12pt]{iopart}
\usepackage{hyperref}

\expandafter\let\csname equation*\endcsname\relax
\expandafter\let\csname endequation*\endcsname\relax
\usepackage{amsmath}

\usepackage{amssymb}
\usepackage[pdftex]{graphicx}

\begin{document}
\DeclareGraphicsExtensions{.pdf}

\title{Emergent gauge dynamics of highly frustrated magnets}
\author{Michael J. Lawler$^{1,2}$}
\address{$^1$ Department of Physics, Applied Physics and Astronomy, Binghamton University, Vestal, NY 13850}
\address{$^2$ Department of Physics, Cornell University, Ithaca, NY 14853}
\ead{mlawler@binghamton.edu}
\submitto{\NJP}
\begin{abstract}
Condensed matter exhibits a wide variety of exotic emergent phenomena such as the fractional quantum Hall effect and the low temperature cooperative behavior of highly frustrated magnets. I consider the classical Hamiltonian dynamics of spins of the latter phenomena using a method introduced by Dirac in the 1950s by assuming they are constrained to their lowest energy configurations as a simplifying measure. Focusing on the kagome antiferromagnet as an example, I find it is a gauge system with topological dynamics and non-locally connected edge states for certain open boundary conditions similar to doubled Chern-Simons electrodynamics expected of a $Z_2$ spin liquid. These dynamics are also similar to electrons in the fractional quantum Hall effect. The classical theory presented here is a first step towards a controlled semi-classical description of the spin liquid phases of many pyrochlore and kagome antiferromagnets and towards a
description of the low energy classical dynamics of the corresponding unconstrained Heisenberg models.
\end{abstract}

\maketitle

\section{Introduction}
There is now substantial experimental\cite{Helton2007, Okamoto2007} and numerical\cite{Leung1993,Lauchli2011,Yan2011} evidence that a combination of frustration and low spin moments is the mechanism that produces novel quantum spin liquid phases in highly frustrated magnets (HFMs). It is not obvious, however, why this should be true. On the kagome lattice, large-$N$ spin models can exhibit order-by-disorder by quantum fluctuations at leading order in the semi-classical large $S/N$ limit\cite{Sachdev1992}. This induced order relieves the frustration of these ``spins'' and places them on similar grounds with unfrustrated ones at smaller $S/N$. So, given the evidence for disordered spin liquid phases, the order-by-disorder phenomena must be much more delicate for ordinary SU(2) spins. Such a conclusion is further backed up by heroic efforts to compute high order large-S expansions\cite{Chubukov1992,Henley1995}.  To address the mechanism that produces these new spin liquid phases then we need a description of these materials that does not begin by relieving the frustration and/or relying on the dominance of order-by-disorder.

To understand the connection between frustration understood at the classical level and the novel phases observed at low spin moments, one way to proceed would be to construct a semi-classical approximation that is still capable of describing non-ordered phases at smaller $S$.   Such a description would be similar to the quantum melting of an antiferromagnetic phase due to order parameter fluctuations\cite{Chakravarty1988} only the dimension of the ``order parameter'' manifold, which are the classical ground states, would be larger, even growing with the size of the system in the highly frustrated case. At present, we know some features of this description. Similar to the distinctions between even, odd and half odd integer spins\cite{Haldane1988,Read1989} in square lattice antiferromagnets, tunneling processes between classical ground states of kagome antiferromagnets leads to distinctions between integer and half odd integer spins\cite{vonDelft1992}. However, it is unclear at present how these tunneling processes and other fluctuations produce a spin liquid phase at smaller spin moments.

To make progress on such a semi-classical description of HFMs, here I study, using Dirac's ``Generalized Hamiltonian mechanics''\cite{Dirac1950,Dirac1958,Weinberg2005,Henneaux1992}, the ``dynamics'' of spins constrained to their ground state configurations. By counting the number of canonical coordinates needed to parametrize this surface, I show that it is a null surface for the Poisson bracket much like the light cone is a null surface in Minkowsky spacetime (See Ch. 2 of \cite{Henneaux1992}). In simpler terms, this means that the number of canonical coordinates of the constrained phase space is not equal to its dimension. Some coordinates are redundant non-canonical coordiantes. According to the Dirac method, this implies the constrained spin model has gauge dynamics. In particular, for the case of periodic boundary conditions, I find that the surface whose dimension is well known to grow with the system size\cite{Chalker1992} has no canonical coordinates. For open boundary conditions, I find that this number depends on the existence of ``dangling triangles'' on the boundary of the kagome cluster and at best grows with its circumference.  The constrained spin kagome model is therefore much like a ``doubled'' version of topological Chern-Simons electrodynamics characteristic of a $Z_2$ spin liquid state\cite{Xu2009}. However, it does not require longer ranged interactions to construct as in the Levin and Wen models\cite{Levin2005}. This approach generalizes easily to other HFMs though the counting of canonical coordinates is likely different in other cases with, for example, the number likely growing with the volume of the system on the pyrochlore lattice. We will conclude with a discussion of the implication of these results for the ordinary unconstrained Heisenberg model and for more realistic models subject to various perturbations.

\section{Kagome ground state spin configurations}
\label{sec:spinconfig}
\begin{figure}[t]
\includegraphics[width=0.33\textwidth]{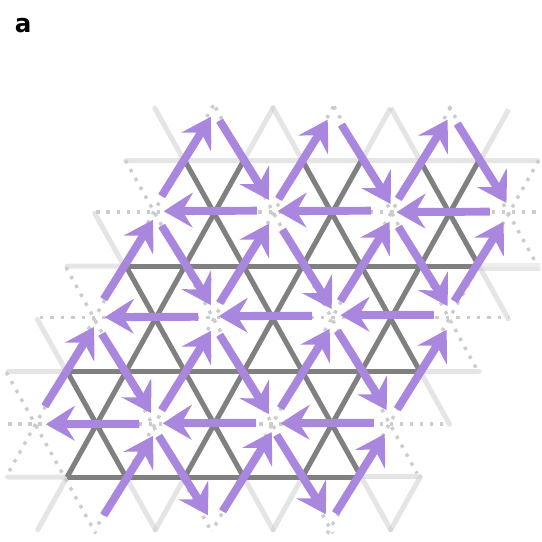}
\includegraphics[width=0.33\textwidth]{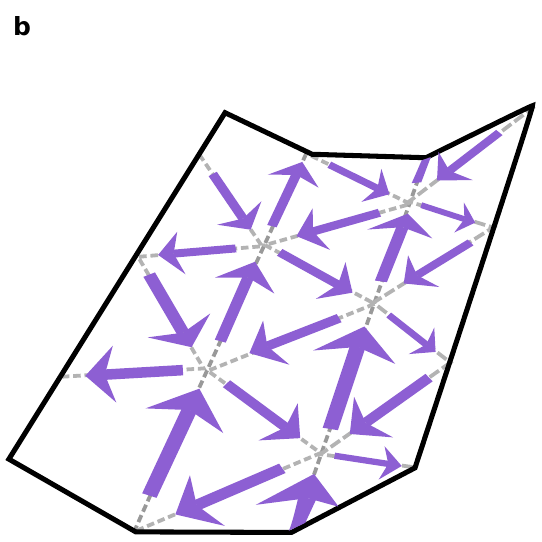}
\includegraphics[width=0.33\textwidth]{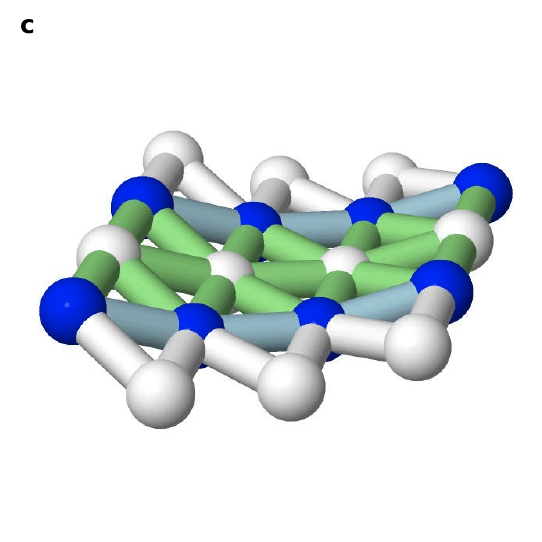}
\caption{The continuously many spin configurations of classical kagome antiferromagnet that have vanishing total spin on each solid triangle. The solid bonds are the kagome lattice, the dashed bonds the triangular lattice of hexagons. {\bf(a)} The coplanar ``$q=0$'' configuration viewed as arrows pointing to vertices of the dashed triangular lattice.  {\bf(b)} A ``folding'' of the spins along a dashed line of the triangular lattice viewed as a piece of paper, called a weather-vane mode\cite{Chalker1992,Ritchey1993, Shender1993}.  The folded spin directions also satisfies \eref{eq:const}. All ground states modes arise from such folding of the spin paper, a construction called ``spin origami''\cite{Ritchey1993}. They imply that low energy spin configurations evolve continuously and collectively in kagome antiferromagnets. {\bf(c)} A visualization of the folded sheet in {\bf (b)} using the solid analogy discussed in the main text.}
\label{fig:kag}
\end{figure}

In a highly frustrated magnet(HFM), spins are frustrated because they have many options to choose from and are unable to decide which is best. On the kagome lattice shown in Fig. \ref{fig:kag}, the classical ground states of the nearest neighbor Heisenberg model prefer a vanishing total spin on each triangle\cite{Chalker1992,Moessner1998a}
\begin{equation}\label{eq:const}
   \phi_{ijk,a} \equiv \Omega^a_i + \Omega^a_j + \Omega^a_k = 0,\quad a\in\{x,y,z\},
\end{equation}
where $\Omega^a_i$, $a\in\{x,y,z\}$ are the three components of the classical spin unit vector $\hat\Omega_i$ on site $i$ and the three sites $i$, $j$ and $k$ form any triangle on the lattice. This may happen in many materials including Herbersmithite, the Jarosite family, SrCr$_{8-x}$G$_{4+x}$O$_{19}$ and Na$_4$Ir$_3$O$_8$. Spins ``suffering'' this condition are highly frustrated for they have difficulty deciding between the continuously many arrangements that satisfy it. Such arrangements are described by the ``spin origami'' construction\cite{Ritchey1993, Shender1993} of drawing spin vectors on a piece of paper and literally folding the paper to obtain new spin directions (see below). The resulting behavior of the spins is then collective and at finite temperatures, they enter a ``cooperative'' paramagnetic phase\cite{Villain1979}. Furthermore, there is a wide class of other HFMs with similar constraints\cite{Moessner1998b} such as the pyrochlore antiferromagnets where the analog of the spin origami construction leads to an effective Maxwell-like gauge description and dipolar spin correlations\cite{Isakov2004,Conlon2009}. Because of their novel low energy properties due to such constrained mechanical behavior, HFMs continue to be promising materials\cite{Gredan2001} to search for new phases of matter.

In essence the origami sheet construction gives us an intuitive representation of the kagome ground state spin configurations. It is a duality transformation to a set of height vectors $\vec h_I$ that live on the triangular lattice formed by the hexagons of the kagome lattice. These vectors are defined through the single spin shared by two neighboring hexagons and are directly analogous to displacement vectors of a two-dimensional solid membrane in three-dimensional space. Specifically, the mapping between height vectors and spin vectors is $\hat\Omega_i = \vec h_I - \vec h_J$ where $IJ$ are the two hexagons that share the spin $i$. This naturally satisfies the constraint that $\hat\Omega_i+\hat\Omega_j+\hat\Omega_k = 0$ on every triangle $ijk$ of the kagome lattice provided a suitable convention for the signs of $\vec h_l$ is made. Because $\hat\Omega_i$ is a unit vector, however, we must always have $||\vec h_I - \vec h_J|| = 1$.  A natural sign convention is to use the spin vectors in the $q=0$ configuration shown in Fig. \ref{fig:kag}(a) where the $\vec h_l$ vector at the tip of a spin arrow enters with a positive sign and the $\vec h_J$  vector at base of the arrow enters with a negative sign. Then setting the height of one of the hexagons to zero (say $\vec h_1=(0,0,0)$), all other heights can then be constructed recursively from the spin vectors such as $\vec h_2 = \vec h_1 + \hat\Omega_1$, etc. The allowed height vectors are then obtained by literally folding the a piece of paper with the $q=0$ spin pattern drawn on it as in \ref{fig:kag}(b). The location of a point on this piece of paper is then $\vec h_l$ and the arrows drawn on the paper are the spin vectors themselves. Fig. \ref{fig:kag}(c) is an alternative presentation of the folded paper in Fig.\ref{fig:kag}(b) viewed as a two-dimensional triangular lattice sheet floating in a three dimensional space where each bond is exactly one unit in length (using Jmol, see \url{http://www.jmol.org}).

\section{Degrees of freedom counting}
\label{sec:dof}
Given the complexity of the kagome ground state spin configurations, an important property to understand are the number of degrees of freedom. One measure of the number of degrees of freedom, discussed in Ref.~\cite{Moessner1998b} and frequently called Maxwell mode counting, is simply the number of free coordinates $d = D -M$ where $D$ is the total number of coordinates and $M$ is the number of constraint functions such as those defined in \eref{eq:const}. Naively, $d=0$ on the kagome lattice because $D=2N$ where $N$ are the number of spins and $M=3N_\Delta$ where $N_\Delta=2N/3$ are the number of triangles. Since we have already constructed several ``folding" modes above proving $d>0$, this naive argument must fail and so the constraints are not all independent (i.e. $M < 3N_\Delta$). 

Given this complexity, a better way to determine $d$ is to look at the number of ways one can fold the spin origami sheet, i.e. its allowed \emph{folding patterns}. The  sheet of Fig. \ref{fig:kag}(b) clearly has two parallel folding modes so $d=2$ for it. We cannot fold along any of the other modes without tearing the piece of paper and violating the constraint $||\vec h_I - \vec h_J|| = 1$. These two folds then define one folding pattern. However, if we were to flatten the paper, we could then fold along one of the other 6 lines.  After this fold, we could no longer fold along the original two lines in Fig. \ref{fig:kag}(b), but would find that we could only fold along two other lines parallel to the one we chose without violating the constraint. Hence, this folding pattern has three allowed folds giving $d=3$. The remaining three lines then make up a third folding pattern.  So the flat sheet here, the coplanar $q=0$ state, is the intersection of three distinct folding patterns. These can alternatively be thought of as three smooth surfaces embedded in the unconstrained $D$ dimensional coordinate space with dimensionalities $d$ taken from $\{2,3,3\}$ that meet at a point that defines the flat sheet. So $d$ takes on different values depending on the origami sheet's folding pattern and the constrained space is a collection of intersecting surfaces defined by these folding patterns in a $D$ dimensional space.

When considering the dynamics of spins, however, the number of free coordinates $d$ is not an important property. Instead it is the number of canonical coordinates $N_c$ that specify the dynamics.  To understand this distinction better, lets demand, as we will throughout this paper, that each spin obey precessional dynamics. We can accomplish this by mapping the azimuthal $\varphi$ and polar $\theta$ coordinates of the spin unit vectors onto a position $q$ and momentum $p$ variable, $\hat\Omega(\varphi,\theta)\to\hat\Omega(q,p)$ such that they obey the usual angular momentum relations $\{\Omega^x,\Omega^y\}=\Omega^z/S$ where $S$ is the spin length or quantum number and $\{,\}$ is the usual Poisson bracket. One choice is $q=\varphi$, $p=S\cos\theta=S\Omega^z$. In an unconstrained spin system, we then see that $N_c = D = 2N$. However, if we were to impose an odd number of constraints $M$ we clearly cannot have $d = N_c$ for $N_c$ must be even. More generally, Dirac found by using Lagrange multipliers to impose the constraints that $N_c <= d$ and that when they are unequal, the extra $N_G = d-N_c$ coordinates are redundant as far as time evolution is concerned and naturally thought of as gauge degrees of freedom. 

\subsection{Triangle and bow-tie models}
\begin{figure}[t]
\includegraphics{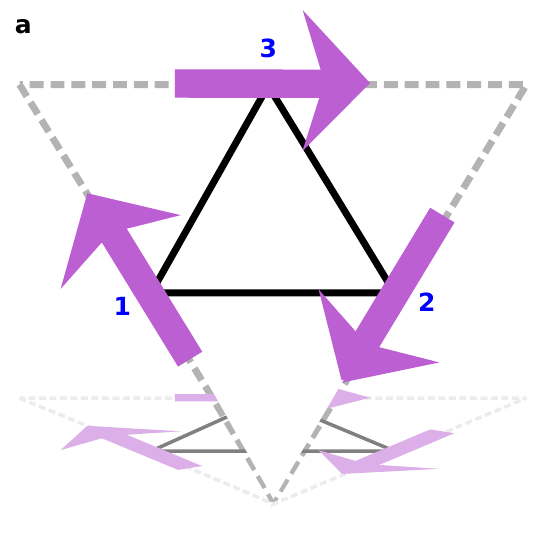}\hspace{1cm}
\includegraphics{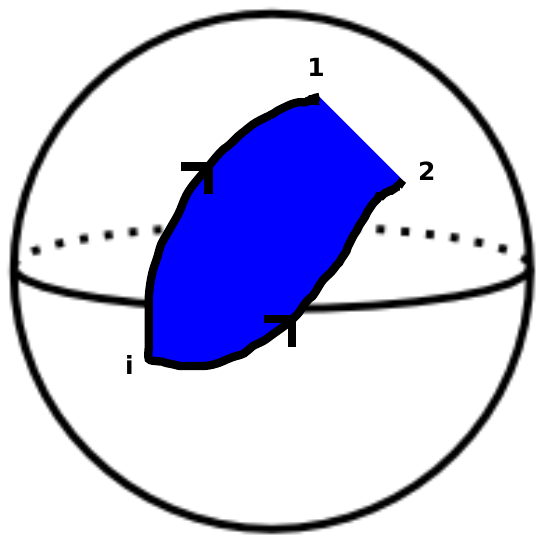}
\caption{Dynamics of the triangle system. {\bf(a)} The three constrained spins of the single triangle that lie on the plane defined by the dashed triangle origami sheet. All spin configurations are then global rotations of the form $\vec S_i = {\bf R}\cdot \vec S_i^0$, that can be viewed as the dashed triangle floating in three dimensional space. {\bf(b)} Two trajectories related by a deformation plotted in the well known solid ball that parameterizes ${\bf R}\in SO(3)$. For each choice of $\vec h_{123}(t)$, a different trajectory will result from solving the equations of motion starting from a given initial condition labeled by {\bf i}. If two choices differ by a smooth function, then we can view one trajectory as a deformation of the other much like a coffee mug and a doughnut can be deformed into each other. Since all trajectories are deformable into each other through these different choices of $\vec h_{123}(t)$, this system has only one possible trajectory.}
\label{fig:Triangle}
\end{figure}

\begin{figure}[t]
\includegraphics{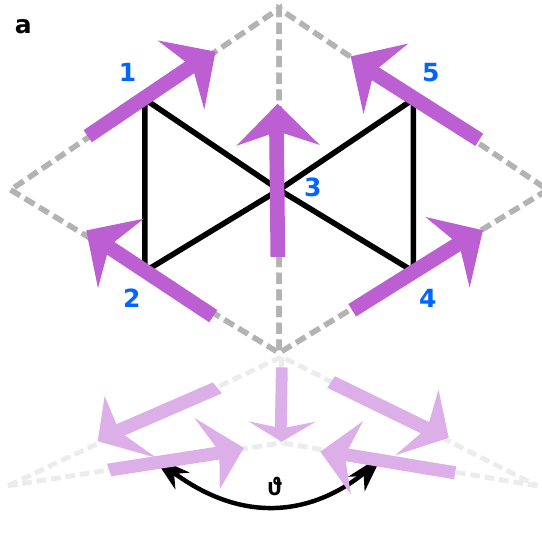}\hspace{1cm}
\includegraphics{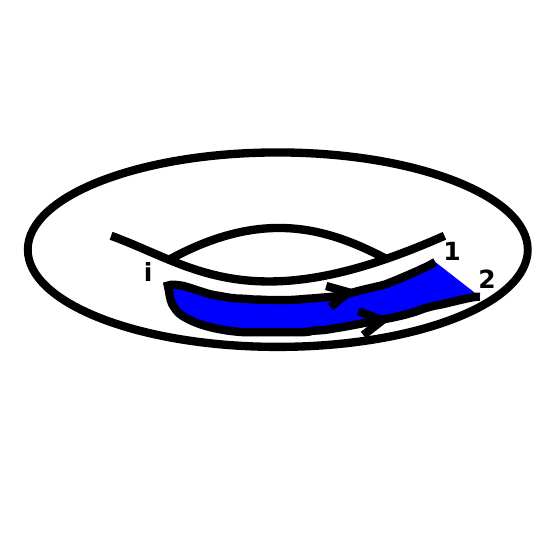}
\caption{Dynamics of the constrained bow tie spin system. {\bf (a)} Origami construction of the spins in the bow tie system viewed as spins drawn on the dashed triangles. All spin configurations are spanned by a choice of $\hat\Omega_3$, the folding angle $\theta$ and an angle $\varphi$ defined by global spin rotations around $\hat\Omega_3$. {\bf (b)} Trajectories in the four dimensional constrained phase space of the bow tie system projected onto the torus formed by the folding angle $\theta$ and the angle $\varphi$ starting from a reference spin configuration. Under time evolution, only the two (gauge) coordinates $\theta$ and $\varphi$ depend on the different choices of $h_{123}(t)$ and $h_{345}(t)$. Any two trajectories with the same $\hat\Omega_3(t)$, such as the two shown, are deformable into each other by changing $h_{123}(t)$ and $h_{345}(t)$.}
\label{fig:BowTie}
\end{figure}

To place the above discussion in a simple context relevant to the dynamics of kagome antiferromagnets, let us find $N_{c}$ for the simpler triangle and bow tie systems, shown in Figs. \ref{fig:Triangle}a and \ref{fig:BowTie}a, systems with only one folding pattern, before turning to the full kagome lattice system. Additional calculations of this sort are provided in \ref{ap:examples}, where $N_c$ is calculated for several well known gauge systems, to place these sort of calculations in a more familiar context.

The first step to determining $N_c$ is to find $d$. The unconstrained phase space of the triangle system, that describes the configurations of the three spins $\hat\Omega(q^1,p_1)$, $\hat\Omega(q^2,p_2)$ and $\hat\Omega(q^3,p_3)$, clearly has $D=6$. Imposing the $M=3$ constraints $\phi_{123,x}=\phi_{123,y}=\phi_{123,z}=0$ then tells us $d=3$ since they are independent constraints. Because $d$ is odd, there is necessarily some ambiguity in identifying canonical degrees of freedom and so this system must have the gauge dynamics discussed above, its phase space must involve at least one gauge coordinate.  

The next step to determining $N_{c}$ it to then work in the unconstrained phase space by introducing Lagrange multipliers and use them to impose the constraints. We do so by extending the Hamiltonian to $H_E = H - \sum_a h_{123}^a\phi_{123,a}$ where the Lagrange multipliers take the form of magnetic fields $\vec h_{123}$. We then choose $\vec h_{123}$ so that $\dot\phi_{123,a}=0$ so that if we obey the constraints at time $t=0$, we will do so for all future times. For a Heisenberg model $H = J\sum_{\langle ij\rangle}\hat\Omega_i\cdot\hat\Omega_j$, I obtain 
\begin{equation}
  \dot\phi_{123,x} = \{\phi_{123,x},H\} - \sum_a h_{123}^a\{\phi_{123,x},\phi_{123,a}\} = 0
\end{equation} 
independent of $\vec h_{123}$ provided initially $\phi_{123,a}=0$ . Similarly $\dot\phi_{123,y} = \dot\phi_{123,z} = 0$ is independent of $\vec h_{123}$.  So there are  $N_L=3$ Lagrange multiplier functions $\vec h_{123}$ undetermined by imposing $\dot\phi_{123,a}=0$ at $t=0$. For each choice of $\vec h_{123}$, we will then get a different time evolution or trajectory for $\hat\Omega_i$ that nevertheless obeys the constraints $\phi_{123,a}=0$! 

The final step then is to make sense of the above result. 
Given a choice of the arbitrary Lagrange multipliers functions $\vec h_{123}(t)$, viewed as a ``choice of gauge", a trajectory in this phase space follows from the equations of motion such as the cartoon picture of \ref{fig:Triangle}a. If we smoothly change $\vec h_{123}(t)$ we would obtain a new ``gauge equivalent" trajectory that can be viewed as a deformation of the first as in  \ref{fig:Triangle}b. The study of the topology of a set works in a similar way. To see that a coffee mug and a doughnut are topologically equivalent we deform the mug into the doughnut. Here, utilizing different choices of $\vec h_{123}(t)$ we can deform trajectories into other trajectories. Since there are $N_L=3$ of these Lagrange multiplier functions the space of trajectories we obtain is necessarily three dimensional and spanned by $N_G\equiv N_L=3$ gauge coordinates. Hence we obtain Dirac's formula for the counting of canonical degrees of freedom: $N_c = d - N_L $ and obtain $N_c=0$ for the triangle system that reflects the absence of any freedom to have different trajectories that are not related by different gauge choices of $\vec h_{123}(t)$. 

Other examples of systems with similar dynamics include a charged particle in a very large magnetic field with a fixed angular momentum\cite{Dunne1990} and ``Chern-Simons'' electrodynamics in two spatial dimensions\cite{Witten1989} (both discussed in \ref{ap:examples}). The latter example is perhaps the most well known for it is related to both Einstein gravity in two spatial dimensions\cite{Witten1988} and the fractional quantum hall effect (FQHE) of a two dimensional electron gas in a very large magnetic field\cite{Girvin1987, Zhang1989,Wen1990b}. Though Chern-Simons electrodynamics has no canonical coordinates like the triangle model, it does have a form of non-trivial dynamics.

In the bow-tie system of two triangles ($M=6$) and five spins ($D=10$, $d=4$), shown in Fig. \ref{fig:BowTie}, the counting of arbitrary Lagrange multipliers demonstrates that it has two canonical and two gauge coordinates. To see this, we set $\dot\phi_{123,a} = \dot\phi_{345,a}=0$ and solve for $\vec h_{123}(t)$ and $\vec h_{345}(t)$. Since any choice with fields parallel to $\hat\Omega_3$, i.e. $\vec h_{123}(t) =  h_{123}(t)\hat\Omega_3$ and $\vec h_{345}(t)=h_{345}(t)\hat\Omega_3$, satisfies these equations, there are two arbitrary Lagrange multiplier functions giving $N_{L}=2$ and hence $N_{c}=2$. Also, by inspection, we can identify the stated two canonical degrees of freedom. These are the two coordinates needed to specify $\hat\Omega_3$ because they evolve in time independent of the choice of lengths $h_{123}$ and $h_{345}$.  After fixing $\hat\Omega_3$, two additional angles are required to specify a configuration of the bow tie system: the ``folding angle'' $\theta$ shown in Fig. \ref{fig:BowTie}a and an angle $\varphi$ that determines the global rotation about $\hat\Omega_3$ of all spins from a chosen reference configuration. As in the triangle model, we can again make a choice of $h_{123}(t)$ and $h_{345}(t)$ and compute a trajectory from the equations of motion. Here we find, by smoothly changing these two functions, that a two dimensional space of trajectories can be deformed into each other all with the same value of $\hat\Omega_3$. Namely, $\hat\Omega_3$ is gauge invariant and its two canonical coordinates span the space of trajectories not deformable into each other through changes in $h_{123}(t)$ and $h_{345}(t)$. Hence, the dynamics is completely specified by the behavior of $\hat\Omega_3$ and this five spin constrained system reduces to the description of a single spin.

\section{Full kagome lattice system}
Consider now the full kagome lattice system. To count and study its canonical degrees of freedom, I have implemented a computational scheme based on the local properties near one spin configuration in the constrained phase space as discussed below. A selection of the counting results are then presented in table \ref{tab:Results}. For periodic boundary conditions, I find $N_{c}=0$ for any sized system, including the smallest system size that is equivalent to and in agreement with the single triangle system discussed above. However, given that the constrained space of the Full kagome lattice is a collection of intersecting surfaces as discussed in \ref{sec:dof}, it is not obvious whether all trajectories can be deformed into each other like in the triangle model. Hence the dynamics of the constrained spins in the kagome antiferromagnet with periodic boundary conditions is composed of one or more discrete sets of topologically equivilant trajectores.

\subsection{Counting Algorithm}\label{subsec:counting}
To study the degrees of freedom of the full kagome lattice model, let us impose \eref{eq:const} on every triangle $\langle ijk\rangle$ of the lattice. To determine the number of arbitrary Lagrange multiplier functions, we extend the Hamiltonian to $H_E = H - \sum_{\langle ijk\rangle} \vec h_{ijk} \cdot \vec\phi_{ijk}$ and solve $\dot\phi_{ijk,a}=0$ so that if at time $t=0$ $\phi_{ijk,a}=0$, it remains zero for all time. Here, $\dot\phi_{ijk,a}$ is given by the usual Poisson bracket relation for the time evolution of any phase space observable:
\begin{equation}\label{eq:phidot}
  \dot \phi_{ijk,a} =  \{\phi_{ijk,a}, H_E\} = \{\phi_{ijk,a}, H\} - \sum_{\langle lmn\rangle,b} h_{lmn,b} \{\phi_{ijk,a},\phi_{lmn,b}\},
\end{equation}
where $\langle ijk\rangle$ and $\langle lmn\rangle$ denotes a triangle, $a$ and $b$ range through $\{x,y,z\}$ and we have assumed $\phi_{ijk,a}=0$ at $t=0$. For the Heisenberg model 
\begin{equation}
  H = JS^2\sum_{\langle ij\rangle}
          \hat\Omega_i\cdot\hat\Omega_j 
      = \frac{J}{2}\sum_{\langle ijk\rangle}\vec\phi_{ijk}^2+ const.
\end{equation}
we have $\{\phi_{ijk,a},H\} = 0$ for initial conditions with $\phi_{ijk,a}=0$. Viewing $\{\phi_{ijk,a},\phi_{lmn,b}\}$ as a square antisymmetric matrix $C_{\alpha,\beta}$, with $\alpha\leftrightarrow ijk,a$ and $\beta \leftrightarrow lmn,b$, we are then left with the following eigenvalue problem
\begin{equation}
  \sum_{\beta} C_{\alpha,\beta} h_{\beta} =0
\end{equation}
If $C_{\alpha,\beta}$ is invertible, then this equation has a unique solution of $h_{\beta} = 0$ and no arbitrary Lagrange multipliers ($N_L = 0$).  However, if $C_{\alpha,\beta}$ is not invertible, if it has zero eigenvalues, then it has many solutions spanned by the null space of $C_{\alpha,\beta}$.  Suppose we find one solution $h_\alpha = h^{(1)}_\alpha$. By knowing an eigenvector $X_\alpha$ of $C_{\alpha,\beta}$ with zero eigenvalue, $h^{(2)}_\alpha = h^{(1)}_\alpha + x(t)X_\alpha$ is also a solution for any $x(t)$. This function $x(t)$ is then an arbitrary Lagrange multiplier. So to determine $N_c$ we need only determine the number of zero eigenvalues of $C_{\alpha,\beta}$, i.e. the dimension of its null space, for this equals the number of arbitrary Lagrange multipliers $N_L$ and $N_c=D-M-N_L$ with $M$ the number of constraint functions $\phi_{ijk,a}$ and $D=2N$ as before. 

The above describes in a nutshell the algorithm for counting the number of canonical degrees of freedom. However, there are two subtleties that need to be considered before applying it. The first is to construct $C_{\alpha,\beta}$ out of an independent set of constraint functions. Given the form of $H_E$, if some $\phi_{ijk,a}$ could be written as a linear combination of the others, it is redundant and should be removed from the set of constraints. We will find this happens frequently. The second subtlety is to avoid computing $C_{\alpha,\beta}$ at a coplanar spin configuration or other point where several distinct folding patterns meet as discussed in \ref{sec:dof}. This is easily achieved by choosing a generic spin configuration.

To find a linearly independent set of constraint functions, lets define a set of vectors, one for each constraint, by taking the phase space gradient of each constraint function, ${\bf v}_\alpha = (\partial\phi_\alpha/\partial q^1, \partial\phi_\alpha/\partial p_1, \ldots)$ evaluated at some point $(q^1,p_1,\ldots)$ in the constrained phase space.  We then need only chose a linearly independent set of vectors ${\bf w}_{\tilde \alpha}$, $\tilde\alpha = 1\ldots M$ using the Gram-Schmidt procedure to obtain $M$.  The constraint matrix of the independent set is then readily evaluated through, 
\begin{equation}
  C_{\tilde \alpha\tilde \beta} =  w_{\tilde \alpha,1} w_{\tilde \beta,2}-w_{\tilde \alpha,2} w_{\tilde \beta,1} + w_{\tilde \alpha,3} w_{\tilde \beta,4}-w_{\tilde \alpha,4} w_{\tilde \beta,3} + \ldots .
\end{equation}
The number of zero eigenvalues of this matrix is then $N_L$ as discussed above.  

A python script implementing this counting algorithm, available online, is discussed in more detail in \ref{ap:Code}. The results for various system sizes and boundary conditions obtained at multiple generic points on smooth portions of the constrained phase space are presented in \ref{tab:Results}. Remarkably, the results for $N_c$ were always the same independent of which folding pattern the generic point belonged.  So unlike $d$, the dimension of a given folding pattern that changes from one pattern/surface to another, the number of canonical coordinates is always the same.

\begin{table}[t]
	\begin{center}
	\begin{tabular}{ | l | c | c | c | c | c | r | }
    		\hline
    		State & Lattice size & Boundary Conditions & $D$ & 
			$M$ & $N_L$ & $N_{c}$\\ \hline
    		Near $\sqrt{3}\times\sqrt{3}$ & $3\times3$ & periodic & 54 & 49 & 5 & 0\\ \hline
    		Near $q=0$ & $3\times3$ & periodic & 54 & 49 & 5 & 0\\ \hline
    		Near $q=0$ & $4\times4$ & periodic & 96 & 90 & 6 & 0\\ \hline
    		Near $q=0$ & $5\times5$ & periodic & 150 & 143 & 7 & 0\\ \hline
    		Near $\sqrt{3}\times\sqrt{3}$ & $3\times3$ & open & 70 & 53 & 11 & 6\\ \hline
    		Near $q=0$ & $3\times3$ & open & 70 & 54 & 10 & 6\\ \hline
    		Near $q=0$ & $4\times4$ & open & 118 & 96 & 14 & 8\\ \hline
    		Near $q=0$ & $5\times5$ & open & 178 & 150 & 18 & 10\\ \hline
    		Near $q=0$ & $3\times3$ & cylindrical---no $\Delta$ & 60 & 51 & 7 & 2\\ \hline
    		Near $q=0$ & $4\times4$ & cylindrical---no $\Delta$ & 104 & 92 & 10 & 2\\ \hline
    		Near $q=0$ & $5\times5$ & cylindrical---no $\Delta$ & 160 & 145 & 13 & 2\\ \hline
    		Near $q=0$ & $3\times3$ & cylindrical---6 $\Delta$& 66 & 54 & 6 & 6\\ \hline
    		Near $q=0$ & $4\times4$ & cylindrical---8 $\Delta$& 112 & 96 & 8 & 8\\ \hline
    		Near $q=0$ & $5\times5$ & cylindrical---10 $\Delta$& 170 & 150 & 10 & 10\\ \hline
        		\hline
  	\end{tabular}
	\end{center}
\caption{Counting of degrees of freedom near two spin configurations and for various lattice sizes and boundary conditions. Here $D$ is twice the number of spins and the number of unconstrained degrees of freedom, $M$ is the number of independent constraint functions and $N_L$ the number of arbitrary Lagrange multipliers. Note: ``cylindrical---no $\Delta$'' means cylindrical boundary conditions without dangling triangles and ``cylindrical---6 $\Delta$'' means with six dangling triangles.}
\label{tab:Results}
\end{table} 

\begin{figure}[t]
\raggedright
\includegraphics{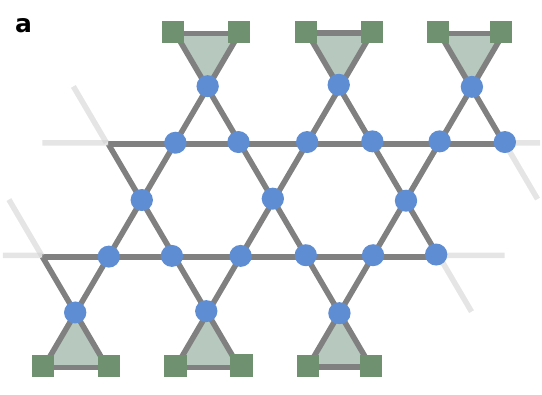}\hspace{1cm}
\includegraphics{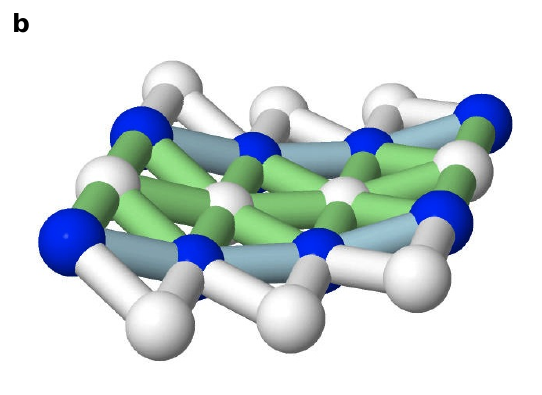}
\includegraphics{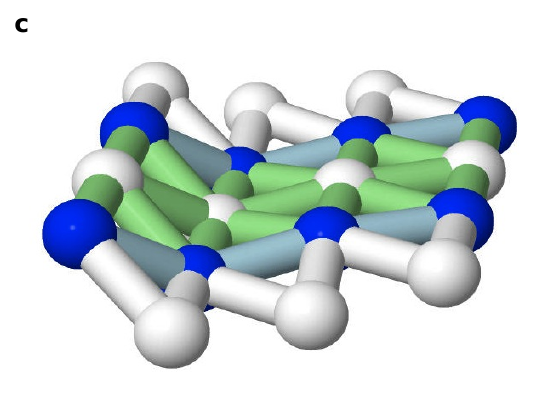}\hspace{1cm}
\includegraphics{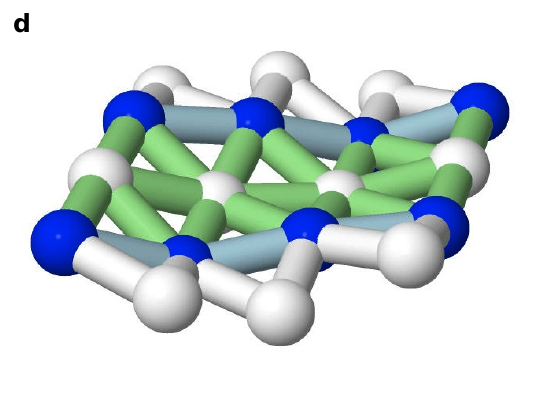}
\caption{Canonical degrees of freedom of the constrained phase space. {\bf (a)} An example of boundary conditions with six dangling triangles highlighted at the top and bottom and $N_{c}=6$. Left and right sides may or may not be connected to form a cylinder. {\bf (b)}-{\bf (d)} Examples of orbits visualized using the origami sheet construction with Jmol (\url{http://www.jmol.org}). The white bonds correspond to the external dangling triangle spins, the blue bonds the third spin on each dangling triangle and the green bonds the bulk spins. Starting from {\bf (b)}, the initial spin configuration, {\bf (c)} is a mode that ``folds'' the left side and {\bf (d)} is the conjugate mode to {\bf (c)} with opposite behavior on the top and bottom blue edge bonds.  Notice how more than one dangling triangle move for both modes, an observation that is generally true of all orbits that changing the canonical degrees of freedom.}
\label{fig:Nc}
\end{figure}

For open boundary conditions, I find $N_{c}>0$ and evidence for unusual edge states. These are similar to the edge-states responsible for the vanishing longitudinal and quantized hall resistance that define the FQHE\cite{Wen1990a}. As shown in table \ref{tab:Results}, $N_{c}$ is equal to the number of ``dangling triangles'' (see Fig. \ref{fig:Nc}(a)) unless there are none in which case $N_{c}=2$. The bow tie system, with two dangling triangles and $N_{c}=2$, is the simplest example of this result. Identifying the canonical degrees of freedom with the dangling triangles leads to a remarkable conclusion. Because two degrees of freedom, a position and a momentum variable, are needed to describe a local mechanical object and there is only one canonical degree of freedom per such triangle, they must be non-locally connected; this result is in agreement with the direct study of the trajectories discussed in the next subsection. 

\subsection{Orbits in the constrained phase space and edges modes}
One way to understand and identify the physical degrees of freedom is to construct the "reduced" phase space of the constrained phase space. That is, if we identify all coordinates that are redundant we can then ignore them and focus on the remaining canonical coordinates. The reduced phase space solves the gauge-redundancy problem explicitly. This solution is particularly useful for our purpose of identifying canonical coordinates numerically.

A nice way to understand the reduced phase space is to construct a basis of tangent vectors at a point $(q^1,p_1,\ldots)$ in the constrained phase space and find the subset of these associated with redundant coordinates. The remaining vectors point along canonical coordinate axes. It turns out that if $X_\alpha$ is an eigenvector of $C_{\alpha,\beta}$ with zero eigenvalue then the phase space vector 
\begin{equation}
   {\bf G} = \sum_{\alpha} X_\alpha \bigg(\frac{\partial \phi_\alpha}{\partial p_1},-\frac{\partial \phi_\alpha}{\partial q^1}, 
    \frac{\partial \phi_\alpha}{\partial p_2},-\frac{\partial \phi_\alpha}{\partial q^2}, \ldots\bigg)
\end{equation}
is tangent to the constrained phase space and points along a gauge coordinate axis known as a ``gauge orbit". The dot product of this vector with phase space gradient of a constraint function ${\bf v}_\alpha$ (defined previously in subsection \ref{subsec:counting}) vanishes for 
\begin{equation}
   {\bf G}\cdot {\bf v}_\alpha = \sum_{\beta} \{\phi_\alpha,\phi_\beta\}X_\beta = \sum_{\beta} C_{\alpha,\beta}X_\beta = 0
\end{equation}
But the vectors ${\bf v}_\alpha$, as gradients of the constraint functions, are a complete set of normal vectors to the constrained phase space surface. So, from the eigenvectors of $C_{\alpha,\beta}$ with zero eigenvalues, we can immediately construct tangent vectors. A complete basis of tangent vectors can then be constructed by finding all vectors ${\bf P}_i$ perpendicular to ${\bf v}_\alpha$ and adding as many of them to the set $\{{\bf G}\}$ as we can to obtain a set $\{{\bf G}_1,{\bf G}_2,\ldots,{\bf P}_1,{\bf P}_2,\ldots\}$ that forms a basis of the tangent space at the point $(q^1,p_1,\ldots)$. 

As stated, the vector ${\bf G}$ points along a gauge orbit; the other ${\bf P}_i$ vectors then point along ``physical'' directions. To see this, consider the equation of motion for any phase space observable
\begin{equation}
  \dot f = \{f,H\} - \sum_\alpha h_\alpha\{f,\phi_\alpha\}
\end{equation}
Letting $h_\alpha = h^{(1)}_\alpha + x(t)X_\alpha$, as discussed in section \ref{subsec:counting}, we obtain
\begin{eqnarray}
  \dot f &= \{f,H\} - \sum_\alpha h^{(1)}_\alpha\{f,\phi_\alpha\} - x(t)\sum_\alpha X_\alpha\{f,\phi_\alpha\}\\
           &= \{f,H\} - \sum_\alpha h^{(1)}_\alpha\{f,\phi_\alpha\} - x(t){\bf F}\cdot{\bf G}\\
           &= \{f,H'\} - x(t) {\bf F}\cdot{\bf G}
\end{eqnarray}
where ${\bf F} = (\partial f/\partial q^1,\partial f/\partial p_1,\ldots)$ and $H' = H - \sum_\alpha h^{(1)}_\alpha \phi_\alpha$.  So ${\bf G}$ is directly associated with the arbitrary Lagrange multiplier function $x(t)$ and any observable $f$ with ${\bf F}$ perpendicular to ${\bf G}$ evolves in time independent of this $x(t)$. Now suppose $f$ was the phase space coordinate $q^1$ and we evolve it in time starting from a given initial conditions $q^1(0)$ under two different choices of $x=x_1$ and $x=x_2$. At an infinitesimal time $t=\varepsilon$ later, the two new points $q^1(t=\varepsilon; x_{1})$ and $q^1(t=\varepsilon; x_{2})$ are related by
\begin{equation}
  q^1(\varepsilon;x_1) - q^1(\varepsilon;x_2) = \epsilon (x_1-x_2) (1,0,\ldots)\cdot {\bf G}
\end{equation}
where $(1,0,\ldots)\cdot{\bf G}$ is just the component of ${\bf G}$ along the $q^1$ direction in phase space. So performing this same calculation for the other coordinates, we see that nearby physically equivalent points in phase space are connected by the vector ${\bf G}$ and path from one of these points to the other is an unphysical gauge orbit. The remaining tangent vectors ${\bf P}_1$, ${\bf P}_2$, $\ldots$ that are linearly independent from ${\bf G}$ then point along directions that cannot be related by a change in $x(t)$. 

Lets now apply this vector analysis to show that trajectories unrelated by a change in $x(t)$ necessarily involve the motion of spins on more than one dangling triangle. To this end, we need to show that all linear combinations of ${\bf P}_1$, ${\bf P}_2$ etc., move spins on more than one dangling triangle. For a given dangling triangle, we then need only define a set of vectors ${\bf  F}=(0,\ldots,1,0,\ldots)$ corresponding to the $q^i$ or $p_i$ coordinate of a bulk spin or a spin on our chosen dangling triangle and show that the set $\{{\bf F}\} + \{{\bf P}\}$ is linearly independent. For the case shown in Fig. \ref{fig:Nc}, the linear independence of this set was found to be true for each dangling triangle. Hence, the canonical modes, as expected from the inability to assign both a position and momentum variable to the spins on a given dangling triangle, are non-local and necessarily involve the motion of spins on more than one dangling triangle.

It is useful to use the spin origami sheet construction to visualize the orbits defined by a tangent vector ${\bf P}_i$. Using this vector, we can construct a small orbit by ${\bf y}(t) = \vec y(0) + t{\bf P}$, $0\leq t\leq\epsilon$. Here ${\bf y} = (q^1,p_1,\ldots)$ is a point in the unconstrained phase space.  This orbit essentially moves a small distance away from ${\bf y}(0)$ and shows how physical modes evolve in time in a given gauge. To see how this affects the spins, it is then straightforward to map ${\bf y}$ to the set of spin vectors $\hat \Omega_i$. It is not easy to understand the collective behavior by directly observing individual spin vectors so lets pass to the spin origami construction for visualization purposes. The result is presented in Fig. \ref{fig:Nc} b-d and movies of them are available online in the supplementary materials. The motion of multiple dangling triangles (that have white or blue bonds) is apparent in both the movie and in Fig. \ref{fig:Nc}(b)-(d). 

\section{Discussion}
In summary using the Dirac approach, I found that the number of canonical coordinates $N_c$, unlike the $d-D-M$ coordinates given by Maxwell counting, is the same on all the intersecting surfaces characterized by folding patterns that make up the constrained phase space. In addition, this $N_c$ vanishes on any cluster with periodic boundary conditions and grows with the number of dangling triangles on clusters with open boundary conditions. The remarkable simplicity of this result allows us to make connections both with known results for the quantum Heisenberg model and for systems whose Hamiltonian is perturbed away from the nearest neighbor model.

The results for $N_c$ here are very similar to doubled Chern-simons electrodynamics expected to describe $Z_2$ spin liquids\cite{Xu2009}. The $Z_2$ spin liquid picture recently gained greater acceptance through DMRG calculations showing a gap to all bulk excitations\cite{Yan2011} and entanglement entropy calculations consistent with a topological $Z_2$ spin liquid\cite{Jiang2012,Depenbrock2012}. This suggests that even spin $1/2$ spins may know about the classical constrained phase space that forms the focus of this paper!

The canonical degree of freedom counting also has implications for realistic Hamiltonians of the form
\begin{equation}
   H = J\sum_{\langle ij\rangle} \vec S_i\cdot\vec S_j + K\left(\text{perturbations}\right)
\end{equation}
where $\vec S_i$ are quantum spin operators, the first term is the nearest neighbor exchange and the second, characterized by an energy scale $K$ that is much smaller than $J$, represents all other perturbations, including further neighbor exchange, Dzyaloshinskii-Moriya interactions, ring exchange and impurities. In the limit $J\to\infty$, the perturbations would induce a Hamiltonian for the constrained phase space with energy scale $K$. Likely this induced Hamiltonian would lead to so-called ``secondary constraints'' and freeze the spins into a single pattern, their ground state. For finite $J$, however, where both gauge and canonical modes are ``physical modes'', the perturbing Hamiltonian would lift the degeneracy of these zero modes giving them dispersion. However, because these modes transform differently under gauge transformations in the constrained model, they should remain as two distinct types of modes. The low energy sector is then spanned by two kinds of zero-modes (the gauge and canonical modes of the constrained phase space) and a third mode involving fluctuations outside of the constrained space such as the ``monopoles'' of Ref. \cite{Conlon2009}. 

\ack
I thank  J. Chalker, C. Henley, E.-A. Kim, S. A.  Kivelson, R. Moessner, P. Nickolic, V. Oganesyan, A. Paramekanti and J. Sethna for useful discussions.

\appendix
\section{Physical degrees of freedom counting for several classical gauge systems}
\label{ap:examples}
\subsection*{Maxwell Electrodynamics}
It is well known that electromagnetic waves are transverse with only two polarizations. Viewed as a Hamiltonian mechanical system, each polarization mode should consist of a position and momentum variable for each point in space. This means that electrodynamics has four canonical degrees of freedom per point in space. However, in the Lagrangian view of electrodyamics (without source terms), the action
\begin{equation}
  S_{Maxwell} = \int d^4x {\mathcal L_{Maxwell}} = -\frac{1}{4}\int d^4x F_{\mu\nu}F^{\mu\nu}
\end{equation}
with $F_{\mu\nu} = \partial_\mu A_\nu - \partial_\nu A_\mu$ is described in terms of four fields $A_\mu = (A_t,\vec A)$, the  scalar and vector potentials. For each ``position'' field $A_\mu$ there should be a ``velocity'' field $\partial_t A_\mu$ making a total of 8 degrees of freedom per point in space. From this perspective, it is not obvious that there are two polarizations of light. Lets show this by passing to the Hamiltonian formalism. A discussion along these lines is also available in Ref. \cite{Weinberg2005} but here I present it in the language of the main body of this paper to facility an understanding of the calculations leading to its main result that $N_c=0$ in the bulk of the kagome constrained spin model. 

To construct a Hamiltonian, we first need to find the momentum fields $\pi^\mu = \delta S_{Maxwell}/\delta \partial_t A_\mu$. They are
\begin{equation}
  \pi^\mu = \partial^\mu A^t - \partial^t A^\mu
\end{equation}
The time component of this equation, $\pi^t\equiv\phi_1=0$ is a phase space constraint, for it does not relate the momentum variable $\pi^t$ to any velocity variables $\partial_tA_\mu$. The spatial components $\vec\pi = -\nabla A_0 - \partial_t \vec A = \vec E$, tells us that the electric field components are the other momentum variables. This means that Gauss's law $\nabla\cdot\vec E\equiv\phi_2=0$ is also a phase space constraint for it is only a relationship between the momentum densities $\vec\pi$. So, the constrained phase space is parametrized by the eight variables $\pi^\mu$ and $A_\mu$ subject to the two constraints $\phi_1=0$ and $\phi_2=0$.

To find the number of physical degrees of freedom per point in space, $N_{c}$, the approach used in the main text is to work in the unconstrained phase space using Lagrange multipliers.  Then $N_{c}$ per point in space is given by $N_{c} = D - M - N_L$ where $D=8$ is the number of unconstrained degrees of freedom, $M=2$ is the number of constraint functions and $N_L$ the number of arbitrary Lagrange multipliers not fixed by the requirement to remain in the constrained phase space. This requirement is constructed by starting from the Hamiltonian, 
\begin{equation}
{\mathcal H} = \int d^3r\big[\pi^\mu \dot A_\mu - {\mathcal L}\big]=\frac{1}{2}\int d^3r\big(\vec\pi^2 + (\nabla\times\vec A)^2\big),
\end{equation}
extending to ${\mathcal H}_E = {\mathcal H} + \int d^3r\big[\lambda_1\phi_1 + \lambda_2\phi_2\big]$ and choosing $\lambda_1$ and $\lambda_2$ so that
\begin{eqnarray}
  \dot\phi_1(\vec r) = \{\phi_1(\vec f),{\mathcal H}\} +\int d^3r' \lambda_2(\vec r')\{\phi_1(\vec r),\phi_2(\vec r')\} = 0,\\ 
  \dot\phi_2(\vec r) = \{\phi_2(\vec r),{\mathcal H}\} + \int d^3r'\lambda_1(\vec r')\{\phi_2(\vec r),\phi_1(\vec r')\} = 0.
\end{eqnarray}
All Poisson brackets in these expressions vanish so both $\lambda_1(\vec r)$ and $\lambda_2(\vec r)$ are arbitrary and $N_L=2$. Hence $N_{c}=4$ per point in space and as discussed above these four degrees of freedom correspond to the two polarizations of light. 

\subsection*{Abelian Chern-Simon's theory}
Abelian Chern-Simon's theory is formally very similar to electrodynamics in two spatial and one time dimension. Maxwell electrodynamics in this two-dimensional case would still have a Lagrangian density $-\frac{1}{4}F_{\mu\nu}F^{\mu\nu}$, just $\mu\in\{t,x,y\}$ does not include the $z$-direction. However, with only two spatial dimensions, the action
\begin{equation}
 S_{Chern-Simons} = \frac{k}{4\pi} \int d^2rdt \varepsilon^{\mu\nu\lambda}A_\mu\partial_\nu A_\lambda
\end{equation} 
is also allowed. Notice it does not involve the ``metric tensor'' $g^{\mu\nu}$ but instead only the antisymmetric tensor $\varepsilon^{\mu\nu\lambda}$. This is a hint that space-time distance, defined by $s^2 = g_{\mu\nu}x^\mu x^\nu = t^2-x^2-y^2-z^2$, may not be important in computing this action but that it may only depend on the topology of the space it is integrating over. It is therefore an interesting alternative to electrodynamics that can arise in a lower dimensional setting. 

To count the physical degrees of freedom of Chern-Simons electrodynamics, let us again pass to the Hamiltonian formalism using the language of the main text. See also Ref. \cite{Dunne1999}.  The unconstrained phase space has $D=6$ degrees of freedom per point in space with the three momentum variables
\begin{equation}
  \pi^t = 0,\quad \pi^x = \frac{k}{2\pi} A_y,\quad \pi^y = -\frac{k}{2\pi} A_x.
\end{equation}
Because none of these three equations involve the velocity variables $\partial_tA_\mu$ they are all phase space constraints. Let us define these constraints through the functions
\begin{equation}
 \phi_t = \pi^t,\quad \phi_x = \pi^x - \frac{k}{2\pi} A_y,\quad \phi_y = \pi^y + \frac{k}{2\pi} A_x
\end{equation}
So we have three constraints $\phi_\mu=0$ from the definition of the momentum variables. To see if there are any other constraints, we need to look at the equations of motion. They turn out to be the three equations
\begin{equation}
  \frac{k}{2\pi}F_{\mu\nu} = \frac{k}{2\pi}\big( \partial_\mu A_\nu - \partial_\nu A_\mu \big)= 0
\end{equation}
so that there are no electric or magnetic fields. One of these equations, $\phi_4 = \frac{k}{2\pi}\big(\partial_xA_y - \partial_yA_x\big)=0$, does not involve any time evolution. It is therefore also a constraint. 

In total then, we have $M=4$ constraints, $\phi_\mu$ and $\phi_4$. Because there are no electric and magnetic fields, the Hamiltonian $H=0$ vanishes. Introducing Lagrange multipliers through 
\begin{equation}
  H_E = H + \int d^2r \big[\lambda^\mu\phi_\mu + \lambda_4\phi_4 \big],
\end{equation}
we need to solve
\begin{equation}
  \dot\phi_a(r) = \sum_b\int d^2r' \{\phi_a(\vec r),\phi_b(\vec r')\}\lambda^b(\vec r') = 
    \sum_b\int d^2r' C_{ab}(\vec r,\vec r')\lambda^b(\vec r') = 0
\end{equation}
for $\lambda^a(\vec r)$ where $b\in\{t,x,y,4\}$. One solution is just $\lambda^a(\vec r)=0$. All solutions define the null space of the constraint ``matrix'' $C_{ab}(\vec r,\vec r')$ and they are spanned by the eigenfunctions of $C_{ab}(\vec r,\vec r')$ with eigenvalue $0$. Computing the matrix explicitly, I find
\begin{equation}\label{eq:CSmatrix}
  {\bf C}(\vec r,\vec r') = \begin{pmatrix} 	0 & 0 & 0 & 0\\
  					0 &  0 & -\frac{k}{\pi}  & \frac{k}{2\pi}\partial'_y \\
					0 & \frac{k}{\pi}  & 0  & -\frac{k}{2\pi}\partial'_x \\
					0 & -\frac{k}{2\pi}\partial'_y  &  \frac{k}{2\pi}\partial'_x  & 0 
		\end{pmatrix}\delta^{(2)}(\vec r-\vec r')
\end{equation}
Fourier transforming and diagonalizing the resulting $4\times4$ matrix, we find two 0 eigenvalues. Transforming back to real space, we discover the null space is spanned by the orthogonal eigenfunctions 
\begin{equation}
 \lambda^t(\vec r) = \psi_1(\vec r), \quad \lambda^x = \lambda^y = \lambda^4 = 0
\end{equation}
and
\begin{equation}
   \lambda^t(\vec r) = 0, \quad \lambda^x(\vec r) = \frac{1}{2}\partial_x\psi_2(\vec r), \quad \lambda^y(\vec r) = \frac{1}{2}\partial_y\psi_2(\vec r), \quad \lambda^4(\vec r) = \psi_2(\vec r),
\end{equation}
where $\psi_1(\vec r)$ and $\psi_2(\vec r)$ are arbitrary functions. These two solutions can be verified by direct substitution into \eref{eq:CSmatrix}. Given the two arbitrary Lagrange multiplier functions $\psi_1(\vec r)$ and $\psi_2(\vec r)$ we have $N_L = 2$ per point in space. Hence, the number of canonical degrees of freedom per point in space is $N_{c} = D - M - N_L = 6 - 4 - 2 = 0$. There are no canonical degrees of freedom in Chern-Simons electrodynamics.

\subsection*{Simple analog of Chern-Simons theory in classical mechanics}
\label{ap:Dunne}
In Ref. \cite{Dunne1990}, {\it ``Topological'' (Chern-Simons) quantum mechanics}, by Dunne, Jackiw and Trugenberger, a simple mechanics model was introduced through an analogy with Chern-Simons gauge theory. Here,  using the language of this paper, I will show how similar in structure this model is to the kagome constrained spin model discussed in the main text.

Ref. \cite{Dunne1990} considers the mechanics of charged particles subject to an external magnetic field in two dimensions (inspired by the fractional quantum hall effect). In the circular gauge, the Lagrangian of such particles is
\begin{equation}
  L = \frac{m}{2} {\bf \dot q}\cdot{\bf\dot q} + eA({\bf q})\cdot {\bf \dot q} = 
    \frac{m}{2} \big(\dot q_x^2 + \dot q_y^2\big) + \frac{eB}{2}\big(q_x\dot q_y - q_y\dot q_x\big)
\end{equation}
In the fractional quantum hall effect, the important limit of the above model is that of very large magnetic fields $B\to\infty$. As such, the authors consider the limit $m\to0$. After taking this limit, the Lagrangian is linear in ${\bf \dot q}$ and the conjugate momenta are
\begin{equation}
  p_x = -\frac{eB}{2} q_y,\quad p_y = \frac{eB}{2}q_x
\end{equation}
So there is no relationship between the momenta and the velocities. In passing to the Hamiltonian formalism, these $M=2$ equations are the constraints
\begin{equation}
  \phi_x = p_x + \frac{eB}{2}q_y,\quad \phi_y = p_y - \frac{eB}{2}q_x.
\end{equation}
in the $D=4$ dimensional unconstrained phase space and the Hamiltonian is $H = p_x\cdot q_x + p_y\cdot q_y - L = 0$. We can work in this phase space by introducing Lagrange multipliers through $H_E = H + u_x\phi_x + u_y\phi_y$ and imposing
\begin{equation}
  \dot \phi_i = \{\phi_i,\phi_j\} u_j = C_{ij} u_j = 0
\end{equation}
where summation over $j$ is implied. The matrix ${\bf C} = eB \left(\begin{smallmatrix} 0 & 1\\-1& 0\end{smallmatrix}\right)$ is invertible so there is one unique solution to this equation $u_i=0$ and there are no arbitrary Lagrange multipliers. This system is not a gauge system. The number of degrees of freedom, all of which are canonical, is $N_{c} = D - M - N_L = 4-2-0=2$.

The focus of the paper, however, is not directly on this model but on the extension of this model to include an additional constraint, that of fixed angular momentum. This alternative model is described by the Lagrangian
\begin{equation}
  L = \frac{eB}{2}\left(q_x\left(\dot q_y - aq_x\right) - q_y\left(\dot q_x + aq_y\right)\right) + \nu a
\end{equation}
where $\nu$ is a parameter and $a$ is a Lagrange multiplier that enforces a third constraint
\begin{equation}
  \phi_J = -\frac{\partial L}{\partial a} = \frac{eB}{2} {\bf q}^2 - \nu = 0
\end{equation}
so that this system has $M=3$. This fixes the angular momentum $J$ because after imposing $\phi_x=0$ and $\phi_y=0$,  $J=q_xp_y - q_yp_x = \frac{eB}{2} {\bf q}^2$.  Following the previous discussion on fixing the Lagrange multipliers, we construct the constraint matrix $C_{ab} = \{ \phi_a,\phi_b\}$, $a = \{x,y,J\}$ and obtain
\begin{equation}
  {\bf C} = \begin{pmatrix} 0 & eB & -eBq_x\\-eB & 0 & -eBq_y\\eBq_x & eBq_y & 0\end{pmatrix}.
\end{equation}
This matrix has eigenvalues $0$, $\pm i\sqrt{B^2 + 2\nu B}$ and the length of the eigenvector that corresponds to the zero eigenvalue
\begin{equation}
  u_x = q_x,\quad u_y = q_y, \quad u_J = 1
\end{equation}
is arbitrary. So this model, with the additional constraint on angular momentum, has one arbitrary Lagrange multiplier and $N_{c} = 4 - 3 - 1 = 0$ physical degrees of freedom. It is very similar in structure to Chern-Simons electrodynamics, but in the context of the mechanics of a particle rather than a field.

\section{Notes on included python scripts}
\label{ap:Code}
There are two python scripts available online. To use them, you need scientific python or pylab. The script ``\verb#state.py#'' holds the information about the spin configurations and lattices and the script ``\verb#constraints.py#'' implements the counting scheme discussed in section 4.1. A typical use of these scripts would be
\begin{verbatim}
In [1]: import state
In [2]: import constraints as con
In [3]: y_cs, T = state.T1()
In [4]: state.check(y_cs,T)
Sum of deviations: 4.4408920985e-16
In [5]: con.report(y_cs,T)
No unconstrained degrees of freedom D:  6
Total number of constraint functions:  3
No of independent constraints M:  3
Dimension d=D-M of the constrained phase spcae:  3
Number of arbitrary Lagrange multipliers N_L:  3
No of physical degrees of freedom:  0
\end{verbatim}
Here \verb#y_cs# is the list of q's and p's associated with a given spin configuration $\{\hat\Omega(q^1,p_1), \hat\Omega(q^2,p_2),\ldots\}$ and \verb#T# is an array that lists the sites in each triangle. In this case, \verb#T# is just \verb#array([[1,2,3]])# because this example is the single triangle system created by the function \verb#state.T1()#. The call to \verb#state.check(y_cs,T)# is to make sure the constraints are obeyed by the spin configuration \verb#y_cs#. To check the bow tie example, use the function \verb#state.T2()# (in this case, \verb#T# is \verb#array([[1,2,3],[3,4,5]])#).

\begin{figure}[t]
\includegraphics[width=0.35\textwidth]{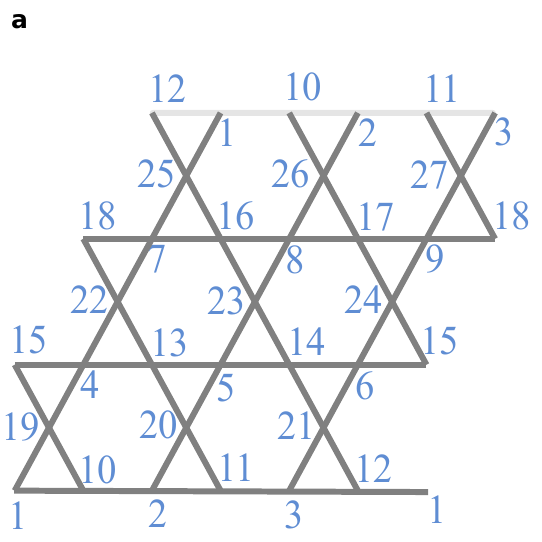}
\includegraphics[width=0.35\textwidth]{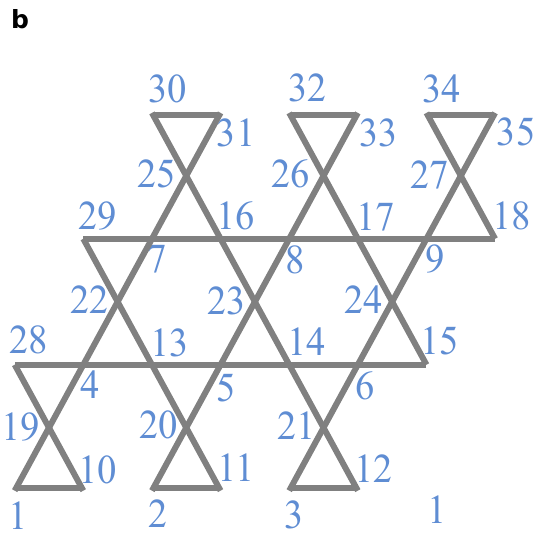}

\vspace{5mm}
\includegraphics[width=0.35\textwidth]{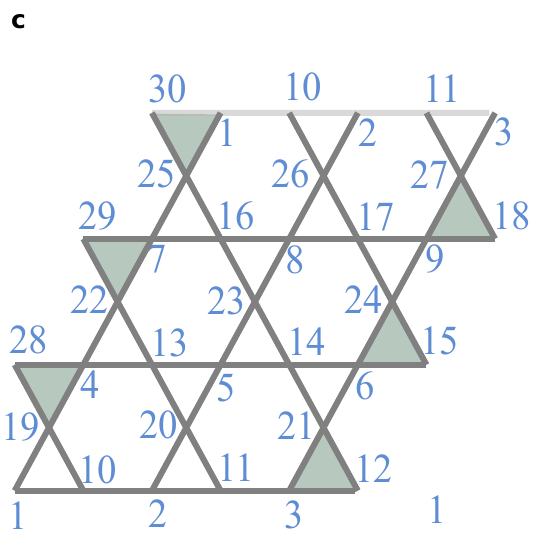}
\includegraphics[width=0.35\textwidth]{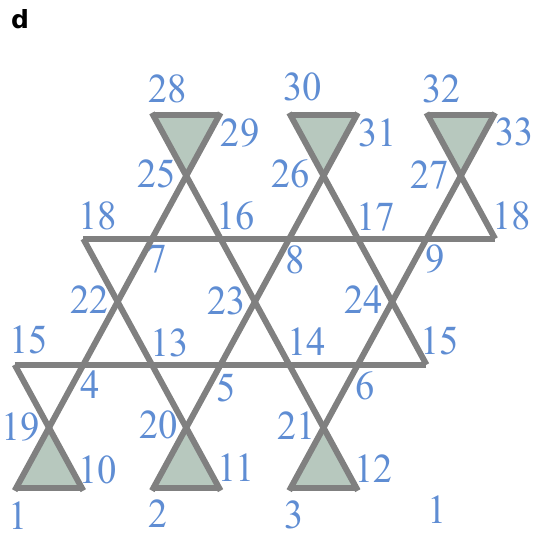}
\caption{Explicit representation of the four boundary conditions used in the code for the $3\times3$ unit cell case. (a) Periodic boundary conditions. (b) Open boundary conditions. (c) cylindrical boundary conditions with no dangling triangles. (d) cylindrical boundary conditions with dangling triangles.}
\label{fig:BCs}
\end{figure}

To study the full kagome lattice, six situations are included. Their boundary conditions for the $3\times3$ unit cell case are explicitly presented in Fig. \ref{fig:BCs}.
\begin{itemize}
\item \verb#state.Kq0pbc(N1,N2,dtheta)#: This function looks at the $N_1\times N_2$ unit cell lattice with periodic boundary conditions where each weather-vane mode is rotated away from the ``$q=0$'' configuration by \verb#dtheta#, 2\verb#dtheta#, 3\verb#dtheta#, etc. If \verb#dtheta# is not specified, then it is set to the small irrational number $\sqrt{2}/100$. 
\item \verb#state.Kq0obc(N1,N2,dtheta)#: This function looks at the $N_1\times N_2$ unit cell lattice with open boundary conditions near the ``$q=0$'' spin configuration.
\item \verb#state.Kq0cbc1(N1,N2,dtheta)#: This function looks at the $N_1\times N_2$ unit cell lattice with cylindrical boundary conditions that leave dangling triangles intact by wrap along the left and right side edges. It is also near the ``$q=0$'' spin configuration.
\item \verb#state.Kq0cbc2(N1,N2,dtheta)#: This function looks at the $N_1\times N_2$ unit cell lattice with cylindrical boundary conditions that leave no dangling triangles by wrapping along the top and bottom leaving the left and right edges free. It is also near the ``$q=0$'' spin configuration.
\item \verb#state.Kr3xr3pbc(dtheta)#: This function looks at the $3\times 3$ unit cell lattice with periodic boundary conditions near the ``$\sqrt{3}\times\sqrt{3}$'' spin configuration.
\item \verb#state.Kq0pbc(N1,N2,dtheta)#: This function looks at the $3\times 3$ unit cell lattice with open boundary conditions near the ``$\sqrt{3}\times\sqrt{3}$'' spin configuration. 
\end{itemize}
These routines in conjunction with \verb#con.report(y_cs,T)# produced the results presented in Table \ref{tab:Results}.


\section*{References}
\begin{thebibliography}{10}

\bibitem{Helton2007}
J.~S. Helton, K.~Matan, M.~P. Shores, E.~A. Nytko, B.~M. Bartlett, Y.~Yoshida,
  Y.~Takano, A.~Suslov, Y.~Qiu, J.-H. Chung, D.~G. Nocera, and Y.~S. Lee.
\newblock Spin dynamics of the spin-1/2 kagome lattice antiferromagnet
  {ZnCu$_3$(OH)$_6$Cl$_2$}.
\newblock {\em Phys. Rev. Lett.}, 98:107204, 2007.

\bibitem{Okamoto2007}
Yoshihiko Okamoto, Minoru Nohara, Hiroko Aruga-Katori, and Hidenori Takagi.
\newblock Spin-liquid state in the s=1/2 hyperkagome antiferromagnet
  {Na$_4$Ir$_3$O$_8$}.
\newblock {\em Phys. Rev. Lett.}, 99:137207, 2007.

\bibitem{Leung1993}
Leung and Elser.
\newblock Numerical studies of a 36-site kagom\'e antiferromagnet.
\newblock {\em Phys. Rev. B}, 47:5459, 1993.

\bibitem{Lauchli2011}
Andres~M. L\"auchli, Julien Sudan, and Erik S.~S\o rensen.
\newblock Ground-state energy and spin gap of spin-1/2 kagom\'e-heisenberg
  antiferromagnetic clusters: Large-scale exact diagonalization results.
\newblock {\em Phys. Rev. B}, 83:212401, 2011.

\bibitem{Yan2011}
Simeng Yan, David~A. Huse, and Steve~R. White.
\newblock Spin-liquid ground state of the s = 1/2 kagome heisenberg
  antiferromagnet.
\newblock {\em Science}, 332:1173, 2011.

\bibitem{Sachdev1992}
S.~Sachdev.
\newblock Kagom\'e- and triangular-lattice heisenberg antiferromagnets:
  Ordering from quantum fluctuations and quantum-disordered ground states with
  unconfined bosonic spinons.
\newblock {\em Phys. Rev. B}, 45:12377, 1992.

\bibitem{Chubukov1992}
A.~Chubukov.
\newblock Order from disorder in a kagom\'e antiferromagnet.
\newblock {\em Phys. Rev. Lett.}, 69:832, 1992.

\bibitem{Henley1995}
C.~L. Henley and E.~P. Chan.
\newblock Ground state selection in a kagom\'e antiferromagnet.
\newblock {\em J. Mag. Mag. Mater.}, 140-144:1693, 1995.

\bibitem{Chakravarty1988}
S.~Chakravarty, B.~I. Halperin, and D.~R. Nelson.
\newblock Low-temperature behavior of two-dimensional quantum antiferromagnets.
\newblock {\em Phys. Rev. Lett.}, 60:1057, 1988.

\bibitem{Haldane1988}
F.~D.~M. Haldane.
\newblock O(3) nonlinear $\sigma$ model and the topological distinction between
  integer- and half-integer-spin antiferromagnets in two dimensions.
\newblock {\em Phys. Rev. Lett.}, 61:1029, 1988.

\bibitem{Read1989}
N.~Read and S.~Sachdev.
\newblock Valence-bond and spin-peierls ground states of low-dimensional
  quantum antiferromagnets.
\newblock {\em Phys. Rev. Lett.}, 62:1694, 1989.

\bibitem{vonDelft1992}
J.~von Delft and C.~L. Henley.
\newblock Destructive quantum interference in spin tunneling problems.
\newblock {\em Phys. Rev. Lett.}, 69:3236, 1992.

\bibitem{Dirac1950}
P.~A.~M. Dirac.
\newblock Generalized hamiltonian dynamics.
\newblock {\em Can. J. of Math.}, 2:129--148, 1950.

\bibitem{Dirac1958}
P.~A.~M. Dirac.
\newblock Generalized hamiltonian dynamics.
\newblock {\em Proc. Roy. Soc. (London)}, A246:326, 1958.

\bibitem{Weinberg2005}
S.~Weinberg.
\newblock {\em The Quantum Theory of Fields: Volume I Foundations}.
\newblock Cambridge University Press, Cambridge, UK, 2005.

\bibitem{Henneaux1992}
M.~Henneaux and C.~Teitelboim.
\newblock {\em Quantization of Gauge Systems}.
\newblock Princeton University Press, Princeton, N.J., 1992.

\bibitem{Chalker1992}
J.~T. Chalker, P.~C.~W. Holdsworth, and E.~F. Shender.
\newblock Hidden order in a frustrated system: properties of the heisenberg
  kagome antiferromagnet.
\newblock {\em Phys. Rev. Lett.}, 68:855, 1992.

\bibitem{Xu2009}
Cenke Xu and Subir Sachdev.
\newblock Global phase diagrams of frustrated quantum antiferromagnets in two
  dimensions: Doubled chern-simons theory.
\newblock {\em Phys. Rev. B}, 79:064405, 2009.

\bibitem{Levin2005}
M.~A. Levin and X.-G. Wen.
\newblock String-net condensation: A physical mechanism for topological phases.
\newblock {\em Phys. Rev. B}, 71:045110, 2005.

\bibitem{Moessner1998a}
R.~Moessner and J.~T. Chalker.
\newblock Properties of a classical spin liquid: The heisenberg pyrochlore
  antiferromagnet.
\newblock {\em Phys. Rev. Lett}, 80:2929, 1998.

\bibitem{Ritchey1993}
I.~Ritchey, P.~Chandra, and P.~Coleman.
\newblock Spin folding in the two-dimensional heisenberg kagom\'e
  antiferromagnet.
\newblock {\em Phys. Rev. B}, 47:15342, 1993.

\bibitem{Shender1993}
E.~F. Shender, V.~B. Cherepanov, P.~C.~W. Holdsworth, and A.~J. Berlinsky.
\newblock Kagome antiferromagnet with defects: Satisfaction, frustration and
  spin folding in a random spin system.
\newblock {\em Phys. Rev. Lett.}, 70:3812, 1993.

\bibitem{Villain1979}
J.~Villain.
\newblock Insulating spin glasses.
\newblock {\em Z. Physik}, 33:31, 1979.

\bibitem{Moessner1998b}
R.~Moessner and J.~T. Chalker.
\newblock Low-temperature properties of classical geometrically frustrated
  antiferromagnets.
\newblock {\em Phys. Rev. B}, 58:12049, 1998.

\bibitem{Isakov2004}
S.~V. Isakov, K.~Gregor, R.~Moessner, and S.~L. Sondhi.
\newblock Dipolar correlations in classical pyrochlore magnets.
\newblock {\em Phys. Rev. Lett.}, 93:167204, 2004.

\bibitem{Conlon2009}
P.~H. Conlon and J.~T. Chalker.
\newblock Spin dynamics in pyrochlore heisenberg antiferromagnets.
\newblock {\em Phys. Rev. Lett.}, 102:237206, 2009.

\bibitem{Gredan2001}
J.~E. Greedan.
\newblock Geometrically frustrated magnetic materials.
\newblock {\em J. Mater. Chem.}, 11:37, 2001.

\bibitem{Dunne1990}
G.~V. Dunne, R.~Jackiw, and C.~A. Trugenberger.
\newblock ``topological'' (chern-simons) quantum mechanics.
\newblock {\em Phys. Rev. D}, 41:661, 1990.

\bibitem{Witten1989}
E.~Witten.
\newblock Quantum field theory and the jones polynomial.
\newblock {\em Commun. Math. Phys.}, 121:351, 1989.

\bibitem{Witten1988}
E.~Witten.
\newblock 2 + 1 dimensional gravity as an exactly soluble system.
\newblock {\em Nuclear Physics B}, 311:46, 1988.

\bibitem{Girvin1987}
S.~M. Girvin and A.~H. MacDonald.
\newblock Off-diagonal long-range order, oblique confinement, and the
  fractional quantum hall effect.
\newblock {\em Phys. Rev. Lett.}, 58:1252, 1987.

\bibitem{Zhang1989}
S.~C. Zhang, T.~H. Hansson, and S.~Kivelson.
\newblock Effective-field-theory model for the fractional quantum hall effect.
\newblock {\em Phys. Rev. Lett.}, 62:82, 1989.

\bibitem{Wen1990b}
X.~G. Wen and Q.~Niu.
\newblock Ground-state degeneracy of the fractional quantum hall states in the
  presence of a random potential and on high-genus riemann surfaces.
\newblock {\em Phys. Rev. B}, 41:9377, 1990.

\bibitem{Wen1990a}
X.~G. Wen.
\newblock Electrodynamical properties of gapless edge excitations in the
  fractional quantum hall states.
\newblock {\em Phys. Rev. Lett.}, 64:2206, 1990.

\bibitem{Jiang2012}
Hong-Chen Jiang, Zhenghan Wang, and Leon Balents.
\newblock Identifying topological order by entanglement entropy.
\newblock arXiv:1205.4289.

\bibitem{Depenbrock2012}
Stefan Depenbrock, Ian~P. McCulloch, and Ulrich Schollwoeck.
\newblock Nature of the spin liquid ground state of the s=1/2 kagome heisenberg
  model.
\newblock arXiv:1205.4858.

\bibitem{Dunne1999}
G.~V. Dunne.
\newblock Aspects of chern-simons theory.
\newblock In {\em Topological aspects of low dimensional systems}, volume~69,
  page 177. Les Houches, 1999.

\end{thebibliography}

\end{document}